\documentclass{IEEEtran}
\usepackage{cite}
\usepackage{amsmath,amssymb,amsfonts}
\usepackage{algorithmic}
\usepackage{graphicx}
\usepackage{bbding}
\usepackage{multirow}
\usepackage{booktabs}
\usepackage[squaren]{SIunits}
\usepackage{longtable}
\usepackage{rotating}
\usepackage{makecell}
\usepackage{pdflscape}
\usepackage{adjustbox}
\usepackage{ulem}
\usepackage{bm,array}
\usepackage[dvipsnames]{xcolor}
\newcommand{\ra}[1]{\renewcommand{\arraystretch}{#1}}

\newcommand{\Rev}[1]{\textcolor{black}{#1}}
\newcolumntype{C}{>{\centering\arraybackslash}p{2em}}
\usepackage{caption}
\usepackage{subcaption}
\usepackage{color,soul}
\newcommand{\vs}[1]{\textcolor{black}{#1}}

\usepackage{subcaption}

\def\BibTeX{{\rm B\kern-.05em{\sc i\kern-.025em b}\kern-.08em
    T\kern-.1667em\lower.7ex\hbox{E}\kern-.125emX}}
\begin{document}

\title{Characterizing the UAV-to-Machine UWB Radio Channel in Smart Factories}


\author{\IEEEauthorblockN{
Vasilii Semkin\IEEEauthorrefmark{1}\IEEEauthorrefmark{2},   
Enrico M. Vitucci\IEEEauthorrefmark{3},   
Franco Fuschini\IEEEauthorrefmark{3},    
Marina Barbiroli\IEEEauthorrefmark{3},      
Vittorio Degli-Esposti\IEEEauthorrefmark{3},      
\\and Claude Oestges\IEEEauthorrefmark{2}      
}                                     

\IEEEauthorblockA{\IEEEauthorrefmark{1}
VTT Technical Research Centre of Finland Ltd., 02150 Espoo, Finland, vasilii.semkin@vtt.fi}\\
\IEEEauthorblockA{\IEEEauthorrefmark{2}
 ICTEAM, Universit\'e catholique de Louvain, 1348 Louvain-la-Neuve, Belgium, claude.oestges@uclouvain.be}\\
\IEEEauthorblockA{\IEEEauthorrefmark{3} 
Department of Electrical, Electronic, and Information Engineering “Guglielmo Marconi” (DEI), \\CNIT, University of Bologna, 40126 Bologna, Italy }

}




\maketitle

\begin{abstract}
In this work, \Rev{the results of Ultra-Wideband air-to-ground measurements carried out in a real-world factory environment are presented and discussed}. With intelligent industrial deployments in mind, we envision a scenario where the \Rev{Unmanned Aerial Vehicle} can be used as a supplementary tool for factory operation, optimization and control. \Rev{Measurements address narrow band and wide band characterization of the wireless radio channel, and can be used for} link budget calculation, interference studies and time dispersion assessment in real factories, without the usual limitation for both radio terminals to be close to ground. The measurements are performed at different locations and different heights over the 3.1-5.3 GHz band. Some fundamental propagation parameters values are determined vs. distance, height and propagation conditions. The measurements are complemented with, and compared to, conventional ground-to-ground measurements with the same setup. The conducted measurement campaign gives an insight for realizing wireless applications in smart connected factories, including UAV-assisted applications.
\end{abstract}

\begin{IEEEkeywords}
UAV, RF channel measurements, UWB radio propagation, smart factory, conscious factory, automation
\end{IEEEkeywords}



\newcommand{\ignore}[1]{}
\newcommand{\inches}{\ensuremath{{}^{\prime\prime}}}
\newcommand{\squeezeup}{\vspace{-6mm}}

\maketitle

\section{Introduction}
\label{sec:introduction}

Nowadays, Unmanned Aerial Vehicles (UAVs), also known as drones, are utilized across a wide range of applications. UAVs introduce new opportunities and increase efficiency in mapping, forensics, visual support for first responders, etc.~\cite{Zeng_UAVcomm_opportunities_2016, Naqvi_drone-aided_2018, Hayat2016_UAVapplications, Kyrkou2019_drones}. An interesting possible application includes UAV utilization in industrial environments. UAVs can facilitate the appearance of new industrial management practices, allowing gathering visual data and performing optimization and control tasks very efficiently, without the need for humans to patrol large, often noisy and sometimes unsafe, industrial premises.

\begin{figure}[!t]
\centering
\includegraphics[scale=0.48]{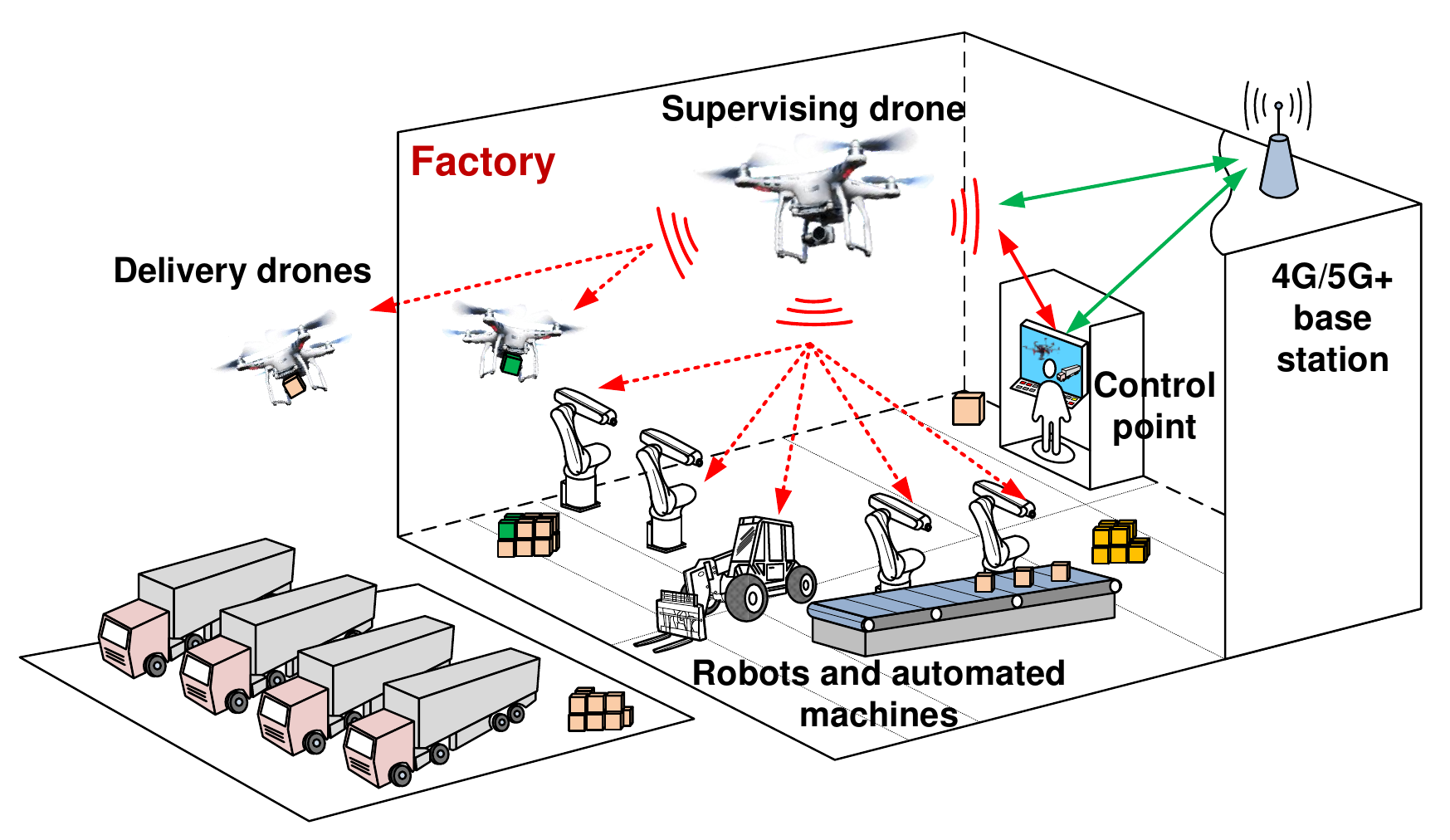}
\caption{Our vision on the UAV integration into the future fully automated conscious factory.}
\label{fig:vision}
\end{figure}

Lately, smart (or "conscious") factories with pervasively connected machines are attracting attention from many industries. The work of different interconnected machines and robots can be optimised and provides economical benefits for companies~\cite{Reimann19}. UAVs, integrated with the other machines can have extensive  capabilities in manufacturing processes, units delivery, supervision, or can even be used to manage and optimize other connected machines, as shown in Fig.~\ref{fig:vision}. The supervising UAV will be operated from the control point. The operator will collect information from the sensors and connected machines about the status of current and future tasks, analyzing factory efficiency and possible problems. If the operator of the factory receives data about defective or inefficient machine operation, the supervising UAV can be sent on place to provide visual data and direct connection between the malfunctioning robot and the operator. This application requires stable wireless connection to the UAV, and the chance that the link is blocked by machines or disturbed by poor propagation conditions must be minimized.
Although the idea of conscious factory was already presented by Nokia and discussed in~\cite{Shu2018_human-robot_interac} in terms of human-robot interaction, to the best authors' knowledge there are no focused studies on radio wave propagation in a factory environment between UAVs and other machines. This is the main reason why we decided to carry out the present work on UAV-to-machine UWB propagation characterization in industrial environment.

\Rev{Ultra-Wide Band (UWB) technology is an established short range communication technology based on the IEEE 802.15.4a/z standard~\cite{IEEE_std_UWB}: numerous applications, including industrial environments, can utilize and benefit from it~\cite{li_timmermann_wiesbeck_zwirello_2013}. UWB radio systems operate in the unlicensed frequency band from 3.1 to 10.6 GHz and offer expedient opportunities for short-range dependable communications at a very low cost. UWB communication systems are of interest also for the large bandwidth available and the low interference due to their low spectral density~\cite{Shen2006_UWB, benedetto2016_uwb}. In addition, UWB communications can be used to serve a large number of users and to achieve a high multi-path resolution, and can be therefore a good candidate for  the rich scattering industrial environment, where high-reliability wireless communication is required. Moreover, UWB transmission can empower accurate radio location that can be very important in smart factory applications~\cite{UWB_UAV2016}.
The exploitation of the UWB high-accuracy localization potential however, as well as the study of propagation in other frequencies bands, e.g. millimeter-wave bands that will be used in next generation wireless systems, are not covered in the present paper but is a part of our future research.}

The communication channel in the UWB frequency band has been thoroughly studied in the past \cite{Karedal2004_UWBfactory, Haneda2006_UWB_chamber, Greenstein2007_UWBcomparison, rusch2003_UWB_resid, ghassemzadeh2002_UWB_in-home, Radovnikovich_UWB_tracking}. In~\cite{Karedal2004_UWBfactory} and \cite{Irahhauten2006_UWB_office+industrial} the authors present Ultra-Wide Band (UWB) channel measurements in an industrial environment along with the analysis of small-scale fading statistics, while in \cite{rusch2003_UWB_resid} and ~\cite{ghassemzadeh2002_UWB_in-home}, the path loss exponent for UWB propagation is evaluated in residential and in-home environments, respectively. \Rev{ In another work presented in~\cite{Tanghe_industrial_meas}, the authors study the industrial indoor channel using fixed positions of the Tx and the Rx at different frequencies, including 5.2 GHz which is a part of the UWB. The obtained path loss exponent values that depend on the environment, e.g. LOS or obstructed LOS (light clutter or heavy clutter) are similar to that one's measured in this work and are discussed in Section~\ref{sec:measresults}.}

\Rev{Although UWB propagation in general has been widely studied, UWB between UAVs and any other machines in industrial environment is barely studied. A few research works study air-to-ground channel properties in indoor environments. In~\cite{Kachroo_UWB2wearables} the authors study communication between UAV and wearable device at UWB frequencies, but the studied environment is an empty warehouse with the metallic walls, therefore very low path loss exponent values are obtained. Most investigations involving UAVs for indoor use are focused on positioning and navigation studies~\cite{li_Tanghe_RFID_localizat, Deng_UAV_camera_tracking}. On the other hand there are many studies related to air-to-ground propagation in outdoor environments, as shown in the following survey~\cite{khawaja_survey}.}

In the present work, we specifically focus on air-to-ground (i.e., UAV-to-robot, UAV-to-operator) UWB radio channel characterization in industrial facilities. Besides their many practical applications, UAVs are also formidable tools for performing measurements at hardly reachable elevated points to give better insight on propagation characteristics in this peculiar environment. Our measurements can be of interests even for cases where the use of a UAV is not necessary, but the knowledge of propagation characteristics for radio terminal located at different heights within an indoor, industrial environment is important for designing reliable communications. In addition, we compare these multiple UAV-to-ground measurements with conventional ground-to-ground measurements in the same environment, to highlight the effect of the UAV platform on the channel characteristics.


\begin{figure}[!t]
\centering
\begin{subfigure}[b]{0.45\textwidth}
{\includegraphics[scale=0.395]{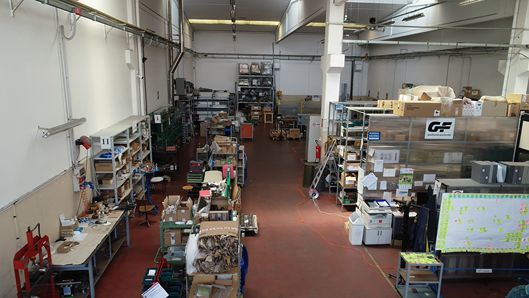}}
\caption{Photograph of the factory}
\end{subfigure}
\begin{subfigure}[b]{0.45\textwidth}
{\includegraphics[scale=0.31]{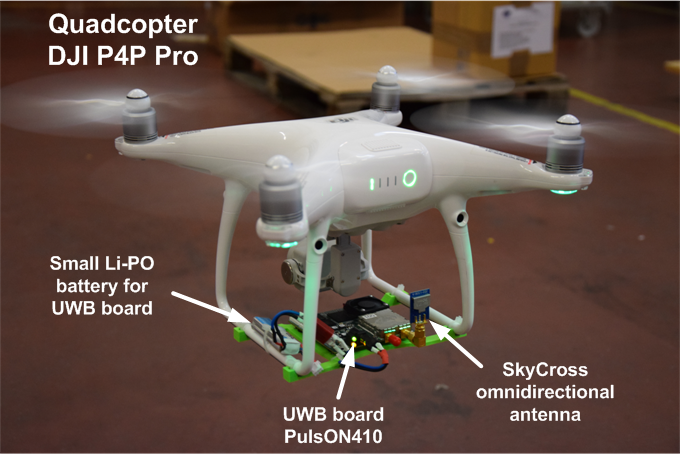}}
\caption{Photograph of the flying UAV}
\end{subfigure}
\caption{Phantom UAV with the transmitting UWB board and the environment of interest.}
\label{fig:drone+factory}
\end{figure}

The structure of this document is as follows: Section~\ref{sec:scenario} presents the measurement setup overview and a description of the measurement scenario. Section~\ref{sec:measresults} provides the analysis of the results. Finally, Section~\ref{sec:conclusion} concludes the paper and discusses plans for the future work.

\section{Measurement equipment and scenario}
\label{sec:scenario}
\begin{figure*}[!t]
\centering
\includegraphics[scale=0.6]{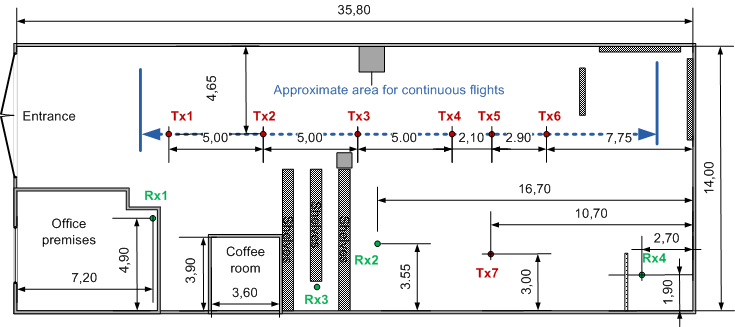}
\caption{Factory floor plan with the dimensions \Rev{in meters} and marked Tx/Rx positions. Green text corresponds to Rx locations, red text to Tx (installed on the UAV) positions.\Rev{The blue dashed-line corresponds to the route of the drone during continuous-flight measurements.}}
\label{fig:factory_plan}
\end{figure*}

The measurements were conducted in a functioning Italian factory (Fig.~\ref{fig:drone+factory}, a) focused on design and construction of customized automation systems. The factory realizes the machinery design and assembly, as well as technical assistance, and administers components storage\footnote{http://www.gfautomazioni.it}. The overall factory's dimensions are 35.8 m by 14 m (the floorplan is presented in Fig.~\ref{fig:factory_plan}). The ceiling height is approximately 7.5 m.


In this measurement campaign, we utilize the popular consumer UAV DJI Phantom 4 Pro (Fig.~\ref{fig:drone+factory}, b). This UAV is able to carry small weight equipment and has multilateral vision system allowing safe indoor manual control. The side obstacle avoiding system was enabled during the UAV operation. It should be noted that UAV was manually controlled during the measurement campaign, since downwards sensor was blocked by the payload (UWB board) and GPS signal was not available indoors.

The measurements were performed using two UWB boards PulsON410\footnote{https://www.humatics.com/} operating from 3.1 to 5.3 GHz. The transmitter (Tx) node was installed under the UAV using a 3D printed fixture. The Tx board was powered with a small Li-Po battery, which is enough for $~$30 min operating time. The receiver (Rx) node was fixed on the mast at a height of 2 m. The particular receiver position Rx1 however is located in the office on the mezzanine floor at 5.4 m height from the ground floor (Fig.~\ref{fig:factory_plan}), to mimic UAV communication with the control room.  Omnidirectional antennas were used at both Tx and Rx sides, with an approximate gain of 3 dBi. Static measurements were performed in most cases (i.e. UAV was hovering at the specified locations marked as Tx[.] in Fig.~\ref{fig:factory_plan}). The channel impulse response (CIR) was recorded for approximately 2 minutes at each location to get the statistics and minimize possible interference from people walking around. During one of the surveying sets the UAV was continuously flying back and forth along the specified route and the CIR was recorded during the whole flight (hereafter called dynamic measurements).

A summary table of the UAV-to-ground measurement results is presented in the Appendix~\ref{app:summary}, Table~\ref{tab:summary} and the locations are specified on the floor plan in Fig.~\ref{fig:factory_plan}. The measurements are organized in Table~\ref{tab:summary} in 3 different data sets, as they have been performed during three campaigns on different dates. Since the factory was operational, there were some differences in the industrial equipment locations. For example, two big metallic cupboards were present only during second measurement set between Tx7 and Rx4 (see Fig.~\ref{fig:factory_plan}). Ground-to-ground measurements (i.e. with the Tx fixed at ground) at several positions were also performed, in order to complement the UAV-to-ground measurements. Also the ground-to-ground measurements were collected in 2 different dates, and they are presented in Appendix~\ref{app:summary}, Table~\ref{tab:summary2}, divided into 2 data sets.

\section{Measurement results}
\label{sec:measresults}
The results are presented and analyzed considering two cases, i.e. static measurements when the UAV is hovering at one of the specified positions and dynamic measurements, when UAV is flying along the factory. In addition, ground-to-ground measurements at several positions are presented: during these measurements the Tx was fixed on the mast at heights of 1, 2, and 3.5 m, respectively. In most cases, the Tx height in ground-to-ground measurements differs from the UAV height in UAV-to-ground measurements in the same locations, except for a few measurements where both the locations and the heights are the same, for the sake of comparison. These additional ground-to-ground measurements are summarized in the Appendix~\ref{app:summary}, Table~\ref{tab:summary2}, and they are compared with the static UAV-to-ground measurements in the following.

\subsection{Measurements at fixed positions}

The measured path loss, RMS delay spread, average and maximum excess delay values for all Tx and Rx locations, \Rev{classified into line-of-sight (LOS), quasi-line-of-sight (quasi-LOS) and non-line-of-sight (NLOS) configurations,} are presented in Table~\ref{tab:summary} and Table~\ref{tab:summary2}.

The mean RMS delay spread values vary between 9.8 and 49.2 ns, where values lower than 25 ns are mainly observed in LOS or quasi-LOS locations, and values higher than 25 ns in NLOS locations, in a reasonable agreement with~\cite{Karedal2004_UWBfactory}. With respect to ~\cite{Karedal2004_UWBfactory} however, lower delay spread values are obtained sometimes for the LOS and quasi-LOS locations: this might be due to different characteristics of the environment. Although the space considered in this work is larger, there is more cluttering due to machines and metallic shelves generating scattering and obstruction. For some locations (e.g. Tx1-Rx1), the UAV-to-ground measurements have been performed twice in different days and the measured delay spread values differs slightly, probably due to the different position of some machines.

\begin{figure}[!h]
\centering
\includegraphics[scale=0.6]{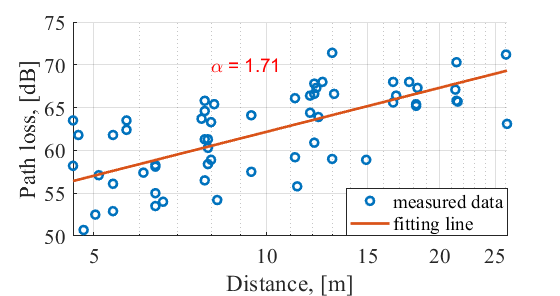}
\caption{Path loss values obtained during all measurements and the corresponding fitting line.}
\label{fig:PL}
\end{figure}

\begin{figure*}[!h]
\centering
\subfloat[RMS Delay Spread]
{\includegraphics[scale=0.55]{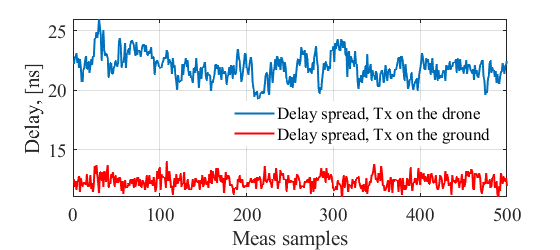}}
\quad
\subfloat[Averaged Power Delay Profile]
{\includegraphics[scale=0.55]{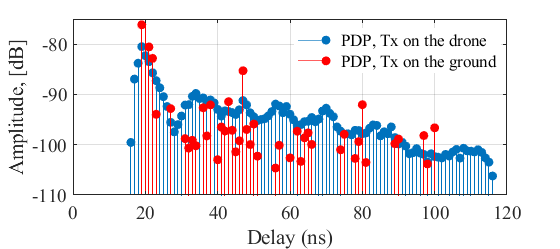}}
\caption{Comparison between UAV and ground measurements for the Tx1-Rx1 setup (blue line - Tx on UAV and red line - Tx on the mast). Rx1 is located upstairs in the office.}
\label{fig:drone_vs_ground}
\end{figure*}

\begin{figure}[!h]
\centering
\subfloat[Average Path Gain vs height.]
{\includegraphics[scale=0.51]{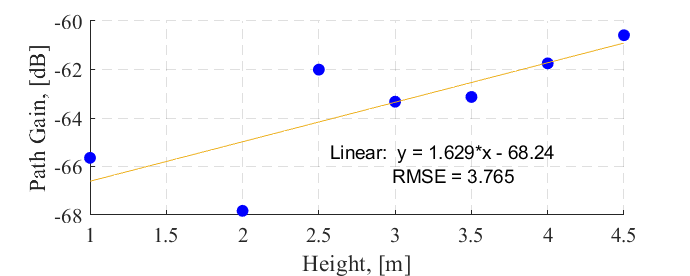}}
\quad
\subfloat[Average RMS Delay Spread vs height.]
{\includegraphics[scale=0.51]{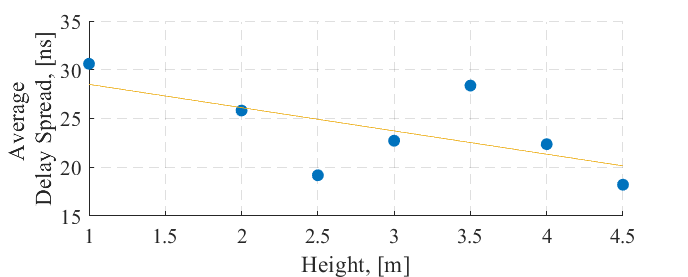}}
\caption{Measured dependence of the average Path Gain and RMS Delay Spread versus height of the transmitter.}
\label{fig:power_and_DS_height_vs_ground}
\vspace{-0.5cm}
\end{figure}

For what concerns path loss, we refer to the well known path-loss exponent model. Specifically, path loss $L(d)$ vs. link-distance is expressed through the following fixed-intercept \Rev{formula~\cite{Rappaport_wireless_comms}}:

\begin{equation}
 L(d) = L(d_0) + 10 \alpha \log_{10}(d) + X_{\sigma},
 \label{eq:logmodel}
\end{equation}
where $d$ is the 3D distance between Tx and Rx, $d_0$ is a reference distance, $\alpha$ is path loss exponent, and $X_{\sigma}$ is a random
variable that accounts for shadowing variation modeled using a zero-mean log-normal distribution with standard deviation $\sigma$, assumed equal to \Rev{the standard deviation of the regression residuals \cite{Rappaport_wireless_comms}}. The data are fitted as usual with a line in a log-log plot, as shown in Fig.~\ref{fig:PL}. Altogether, sixty measurements are selected to determine the path loss values: it is worth noticing that the fast fading effect is removed from the data due to the large bandwidth of the UWB equipment and the 2-minutes time-averaging at each static location.
Path loss values in different locations have large deviations due to the rich scattering environment. Although the factory can be considered mostly an open space, the many metal scatterers generate a quasi-reverberant environment. The estimated path loss exponent is equal to 1.71, i.e. lower than free-space, which is rather common for such a type of environment. This value is similar to that ones found in the literature, \vs{e.g.~\cite{rusch2003_UWB_resid, ghassemzadeh2002_UWB_in-home}}. The corresponding value for the standard deviation of the shadowing is $\sigma=3.8$ dB.

One may observe that the setup when the UAV is hovering replicates static ground measurements when both Tx and Rx are fixed on tripods or masts. The setups are similar, but the UAV suffers from vibrations and some drifting relative to its initial position due to the GPS being blocked by the roof, and the UAV body itself impacts on the multipath structure. As an example, in Fig.~\ref{fig:drone_vs_ground} we compare RMS delay spread and averaged PDP for the "UAV-to-ground" and "ground-to-ground" measurements for Tx1-Rx1, with Tx at 2 meters height in both cases. It is clear that results are different for the two cases. Besides drifting, which is evident from the greater irregularity of the curve in Fig.~\ref{fig:drone_vs_ground}(a), the UAV body is affecting the radiation pattern of the antenna and generating reflections or obstructions from the UAV hull. Therefore, the PDP has a denser structure when compared to the corresponding ground-to-ground case: this also corresponds to a higher average Delay Spread value (22 ns vs. 9.8 ns). A similar behaviour can be observed comparing UAV-to-ground vs. ground-to-ground measurements with same height in other locations (e.g. Rx1-Tx2).

\Rev{However, if we consider the whole measurement dataset in Table~\ref{tab:summary} (UAV-to-ground measurements) and Table~\ref{tab:summary2} (ground-to-ground measurements) of Appendix~\ref{app:summary}, on the average the reported values do not differ much. Considering the average delay spread for all the considered locations, both with Tx on the UAV (Table~\ref{tab:summary}) and Tx on the mast (Table~\ref{tab:summary2}), we get an average delay spread value of 18 ns in the LOS and quasi-LOS configurations, and of 28 ns in the NLOS configurations. Regarding path loss, we get an average value of 59 dB in the LOS/quasi-LOS configurations, and of 66 dB in the NLOS configurations.} \Rev{Nevertheless, the measured values in the different Rx locations are strongly influenced by the degree of obstruction caused by shelves and machines, and especially by the Tx height.}

In Fig.~\ref{fig:power_and_DS_height_vs_ground}, the dependence of the average normalized power (path gain) and RMS delay spread versus Tx height is presented. Only a subset of the measured data (both UAV-to-ground and ground-to-ground) is considered in this plot: the considered data are marked with a "*" in Table~\ref{tab:summary} and Table~\ref{tab:summary2} (see Appendix~\ref{app:summary}), and they correspond to those locations that are more representative of typical configurations (LOS,  quasi-LOS, and NLOS). Moreover, all the Path Gain values have been realigned to remove the dependence on  distance, in order to observe only the effect of Tx height and obstructions.
Looking at Fig.~\ref{fig:power_and_DS_height_vs_ground}(a), it is evident that the path gain is linearly increasing for heights above 2.5~m, while for lower heights the deviations from the fitting line are more severe. Being the height of most shelves of the factory approximately 2.5 m, we believe that these fluctuations for lower heights are probably caused by obstructions caused by objects on the shelves. The average RMS Delay Spread (Fig.~\ref{fig:power_and_DS_height_vs_ground}(b)) also varies significantly with the height according to an overall decreasing trend.

\subsection{Measurements during continuous flight}

In this subsection we present the results obtained during the continuous UAV flight at height of approximately 2.5 m. The starting point is located near the entrance and the UAV flies straight to the opposite wall and back, forward and backward for five times. \Rev{The continuous flight route is represented with a blue dashed line in Fig.~\ref{fig:factory_plan}. The receiver is fixed at position Rx2 (see Fig.~\ref{fig:factory_plan}), at 2 m height}. There might be an error of 0.5-1 m in the UAV heights and also along x- and y- coordinates since the UAV was operated manually and was drifting during flight.

In Fig.~\ref{fig:path_gain+delay_spread_cont}, (blue line) the average measured path gain during continuous UAV flights is shown. \Rev{The path gain is obtained by integrating the CIR for each snapshot, and averaging the obtained values through a sliding observation windows of about 1 m: after that, the obtained path gain values for the 5 round-trip flights are averaged together.} The spatial averaging procedure is necessary to cut-off fast-fading effects and to clearly identify starting points for each flight, so that the subsequent time averaging of the data among the flights becomes more accurate.
\Rev{The speed of the UAV was almost constant for all conducted flights and is equal to approximately 1 m/s. There might be some difference in the beginning and end of the route due to acceleration and braking of the UAV. The averaging procedure previously described can help to compensate for possible different speed at these parts of the route.}\Rev{Besides averaging, the standard deviation of the collected data for each reference point of the route is also computed, and represented through vertical error bars in Fig.~\ref{fig:path_gain+delay_spread_cont} (the length of the vertical bars is equal to double the standard deviation).}

Looking at Fig.~\ref{fig:factory_plan}, we observe that the initial UAV position has no LOS link and moreover, there are many shelves between Tx and Rx. In fact, as the UAV flies towards the end wall, the power level (path gain) shown in Fig.~\ref{fig:path_gain+delay_spread_cont} increases and reaches a maximum for the (quasi)-LOS case when UAV is in the middle of the route at approximately 10m from the starting point, i.e. when the Tx-Rx distance is minimum. Then the power level drops due to the increasing distance between Tx and Rx and some machines obstructing the link, but not as much as for the first part of the route.

In Fig.~\ref{fig:path_gain+delay_spread_cont} (red line), the mean and standard deviation values of the delay spread are also presented, for the forward and backward UAV flights. The measured average delay spread for continuous UAV flight varies from \vs{10 ns to 23 ns}.

Interestingly, the standard deviation of both path loss and  delay spread is greater in the central part of the graph: this is probably due to the quasi-LOS nature of the locations in this section, where obstructing objects can generate intermittent shadowing and therefore a great variability.

\subsection{Comparison with previous works}
\Rev{The work~\cite{Shokouh_2009_UWB_review}, summarizes the results from many UWB indoor measurement campaigns (in office, classroom, laboratory environment, etc.) and shows that the path loss exponent values are in between 1.3 and 2.4 for LOS case, while typical values above 1.7 are found in the NLOS cases. Another work~\cite{Rappaport1989_UHF_fact}, analyses narrowband measurement results obtained in five factories at ultra-high frequencies, i.e. 1.3 GHz: the obtained path loss exponent values are equal to 1.79 for LOS cases (light and heavy clutter). In general, values lower than 2 are quite common in some indoor scenarios (e.g. corridors or large rooms), because the confined environment allows the establishment of waveguiding effects.}

\Rev{According to~\cite{Rappaport_wireless_comms}, the following path loss exponent values are expected in different environments at frequencies below 6 GHz:}
\Rev{\begin{itemize}
    \item{Urban area cellular radio (non-shadowed)=2.7-3.5}
    \item{Shadowed urban cellular radio=3-5}
    \item{In-building LOS=1.6-1.8}
    \item{Obstructed LOS in-building=4-6}
    \item{Obstructed LOS in factories=2-3}
\end{itemize}}
\Rev{
 Based on the information presented above, initially, we expected to get a path loss exponent value from 1.6 up to a maximum of 3. However, due to large open spaces in the factory and high probability of LOS communication, since in most cases the UAV is flying higher than average shelf height, the obtained path loss exponent value is equal to 1.71 which is overall in good agreement with the results of the studies reported above, for similar frequency bands.}

\Rev{The path loss exponent value obtained in this work is also similar to what found in other investigations at UWB frequencies in residential environment \cite{rusch2003_UWB_resid, ghassemzadeh2002_UWB_in-home}: probably, the heavily cluttered industrial environment we considered shows a similar degree of obstruction despite the larger size of the room. In addition, the results are similar to those ones obtained in~\cite{Briso_UWB_indoors}, where the authors study UWB propagation in a large open indoor environment, i.e. a sports hall.}

\Rev{The value of path loss standard deviation due to shadowing is approximately 4 dB in our measurements, which is in good agreement with the literature. For example, in~\cite{benedetto2016_uwb,Tanghe_industrial_meas}, the authors report shadowing standard deviation values of 4 to 5 dB in the 5.2 GHz band, and around 4 dB at 3.9 GHz in~\cite{Briso_UWB_indoors}. The average delay spread values (from 9.8 to 49.2 ns) are also in good agreement with the measurements described in~\cite{benedetto2016_uwb} and~\cite{Karedal2004_UWBfactory}.}




\ignore{
\begin{figure}[!h]
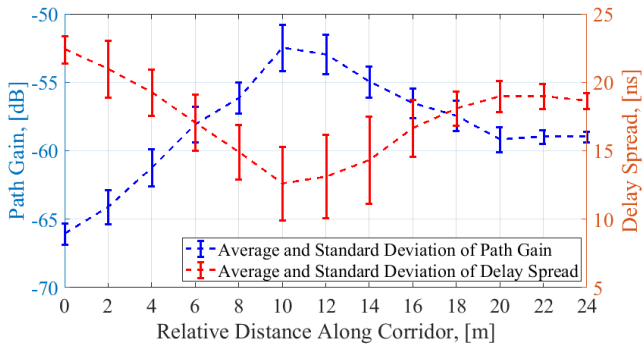

\centering
\subfloat[Path gain during the continuous flight.]
{\includegraphics[scale=0.51]{Images/PathGain_along_corridor.png}}
\quad
\subfloat[Delay spread during the continuous flight.]
{\includegraphics[scale=0.51]{Images/delayspread_along_corridor.png}}
\caption{Average and standard deviation of the path gain and delay spread during continuous UAV flights.}
\label{fig:path_gain+delay_spread_cont}
\end{figure}}

\begin{figure}[!t]
\centering
\includegraphics[scale=0.38]{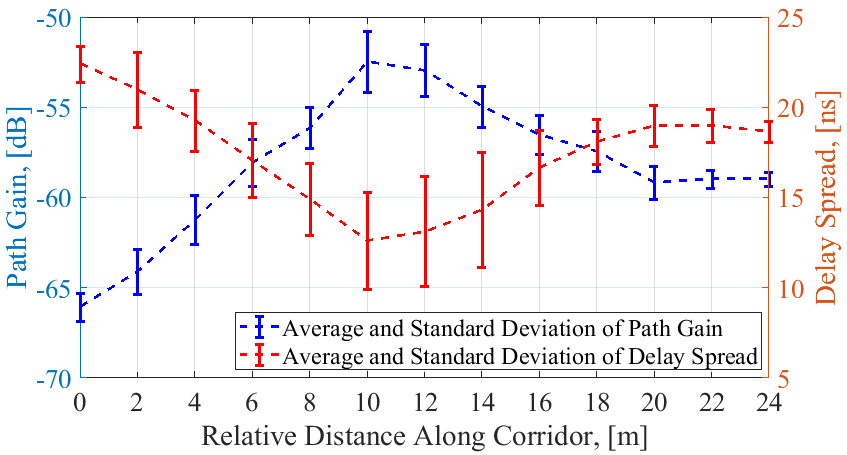}
\caption{Average and standard deviation of the path gain and delay spread during continuous UAV flights.}
\label{fig:path_gain+delay_spread_cont}
\end{figure}

\section{Conclusion}
\label{sec:conclusion}
In this work, we present the results of a measurement campaign and a detailed analysis of air-to-ground radio channel characteristics at UWB frequencies. The measurements were conducted in a real, mid-size industrial environment with Tx on a hovering/flying UAV for most cases. Conducted measurements replicate the realistic case of a supervising UAV which is used at the conscious factory to provide visual control for the operational robots and machines. 

\Rev{The main channel characteristics (path gain, delay spread, path-loss exponent, shadowing standard deviation) are analyzed in detail in different propagation conditions (LOS, quasi-LOS and NLOS) and at different heights, both with the UAV hovering in a fixed position and with the UAV flying continuously along the main corridor of the factory.} The results presented in this work may be used for link budget calculations and interference analysis.
A  comparison of "UAV-to-ground" measurements vs. "ground-to-ground" measurements for the same positions shows differences in the obtained values for some of the considered measurement locations, due to UAV drifting and to the obstruction/scattering effect of the UAV body on the signal.

\Rev{A literature overview of previous work on UWB channel measurements in factories, large-indoor, and air-to-ground environment is presented in the paper together with a discussion of results. The values found for Path loss exponent, shadowing standard deviation, and RMS Delay Spread, are in agreement to those found in previous work for similar environments and frequencies.}

Future work will deal with studies at different frequency bands, including mm-wave bands and with the characterisation of the directional properties of the channel in industrial environment.

\section*{Acknowledgment}
The authors are grateful to GF automazioni for the opportunity to perform measurements in an operational factory, to Prof. Marco Chiani and Prof. Davide Dardari of the University of Bologna for providing the UWB boards, and to our students Roberto Cirillo and Vincenzo Nardiello for assisting us in measurement campaign and data post-processing.



\bibliographystyle{IEEEtran}
\bibliography{refs}



\begin{appendices}
\section{Summary of the performed measurements}
\label{app:summary}

\begin{table*}[t]
    \centering
    \caption{Summary of the performed measurements with Tx on UAV}
    \ra{1}
    \begin{tabular}[font=\small]{|c|c|c|c|c|c|c|c|c|c|}
        \hline
        \makecell{Meas.\\Set } & \makecell{UAV (Tx)\\Pos.} & \makecell{Rx\\Pos.} & \makecell{Scenario\\Type } &\makecell{UAV (Tx)\\Height, [m]} & \makecell{Tx-Rx \\ Dist., [m]} & \makecell{Mean Path \\ Loss, [dB]} & \makecell{Mean Excess\\Delay, [ns]} & \makecell{Max Excess\\Delay, [ns]} & \makecell{RMS Delay \\Spread, [ns]}\\
        \hline
        \multirow{10}{*}{\makecell{N\textsuperscript{\underline{o}}1}\ignore{29.06.2019}} & Tx7 & \multirow{4}{*}{Rx3} & NLOS & 4 & 5.4 & 61.8 & 31.7 & 95.7 & 18.6\\
        & Tx2 &  & NLOS & 2 & 7.9 & 60.3 & 27.8 & 90.2 & 13.7\\
        & Tx1 &  & NLOS & 2 & 12.1 & 60.9 & 29 & 94.4 & 15.1\\
        & Tx1 &  & NLOS & 4 & 12.3 & 63.9 & 31.7 & 95.7 & 17.1\\ \cline{2-10}
        & Tx6 & \multirow{6}{*}{Rx1} & NLOS & 2 & 21.3 & 67.1 & 31.4 & 107.1 & 21.9\\
        & Tx7 * &  & NLOS & 4 & 16.6 & 65.6 & 36.1 & 97.8 & 24.4\\
        & Tx3 * &  & NLOS & 4 & 11.9 & 64.4 & 41.3 & 98.6 & 23.2\\
        & Tx2 * &  & Q-LOS & 4 & 7.7 & 63.7 & 42 & 95.1 & 24.2\\
        & Tx1 * &  & Q-LOS & 2 & 5.7 & 62.4 & 27.2 & 94.2 & 16.7\\
        & Tx1 * &  & Q-LOS & 4 & 4.7 & 61.8 & 32.9 & 95.7 & 20.2\\
        \hline
        \multirow{8}{*}{\makecell{N\textsuperscript{\underline{o}}2}\ignore{18.09.2019}} & Tx1 * & \multirow{8}{*}{Rx2} & NLOS & 2.5 & 12.1 & 67.8 & 34.7 & 100.8 & 20.8\\
        & Tx2 * &  & NLOS & 2.5 & 7.9 & 61.3 & 28.6 & 94.2 & 15.5\\
        & Tx3 &  & Q-LOS & 2.5 & 5.1 & 57.1 & 26.5 & 74.5 & 12.4\\
        & Tx4 * &  & LOS & 2.5 & 6.4 & 58.3 & 27.6 & 91.8 & 15.4\\
        & Tx7 &  & LOS & 2.5 & 6.1 & 57.4 & 30 & 93.2 & 20.8\\
        & Tx5 &  & LOS & 4.5 & 8.2 & 54.2 & 23.5 & 83.1 & 11.2\\
        & Cont. flight &  & --- & 2.5 & 5 - 15 & 53 - 66 & 19.5 - 47.2 & 54 - 114 & 10 - 23\\
        \hline
        \multirow{14}{*}{\makecell{N\textsuperscript{\underline{o}}3}\ignore{19.09.2019}} & Tx1 & \multirow{7}{*}{Rx4} & NLOS & 2 & 26.1 & 71.2 & 33.6 & 114.1 & 24.9\\
        & Tx2 * &  & NLOS & 2.5 & 21.4 & 65.8 & 41.2 & 108.8 & 25\\
        & Tx3 &  & NLOS & 2.25 & 16.8 & 66.4 & 47.7 & 109.8 & 17.7\\
        & Tx4 &  & Q-LOS & 2.5 & 12.5 & 68 & 42 & 112.4 & 14.7\\
        & Tx7 &  & Q-LOS & 3 & 8.1 & 65.4 & 34.6 & 96.3 & 18.3\\
        & Tx1 &  & NLOS & 4.25 & 26.2 & 63.1 & 32.3 & 103.4 & 19\\
        & Tx5 &  & Q-LOS & 4.75 & 11.3 & 55.8 & 27.2 & 86.6 & 13.4\\ \cline{2-10}
        & Tx1 * & \multirow{7}{*}{Rx1} & Q-LOS & 2 & 5.7 & 63.5 & 34.9 & 102.2 & 22\\
        & Tx1 * &  & Q-LOS & 4.5 & 4.6 & 58.2 & 31.2 & 87.2 & 17.4\\
        & Tx2 * &  & Q-LOS & 2.75 & 7.8 & 65.8 & 41.2 & 108.8 & 22.1\\
        & Tx3 &  & NLOS & 2.5 & 11.9 & 66.4 & 47.7 & 109.8 & 24.1\\
        & Tx4 &  & NLOS & 2.5 & 16.6 & 68 & 42 & 112.4 & 26.8\\
        & Tx5 * &  & NLOS & 4.5 & 17.7 & 68 & 43.7 & 112.8 & 26.2\\
        & Tx7 * &  & NLOS & 2.75 & 18.2 & 65.4 & 34.6 & 96.3 & 23.4\\
        \hline
    \end{tabular}
    \label{tab:summary}

\end{table*}{}

\begin{table*}[t]
    \centering
    \caption{Summary of the performed measurements with Tx at ground}
    \ra{1}
    \begin{tabular}[font=\small]{|c|c|c|c|c|c|c|c|c|c|}
        \hline
        \makecell{Meas.\\Set } & \makecell{Tx\\Pos.} & \makecell{Rx\\Pos.} & \makecell{Scenario\\Type } & \makecell{Tx\\Height, [m]} & \makecell{Tx-Rx \\ Dist., [m]} & \makecell{Mean Path \\ Loss, [dB]} & \makecell{Mean Excess\\Delay, [ns]} & \makecell{Max Excess\\Delay, [ns]} & \makecell{RMS Delay \\Spread, [ns]}\\
        \hline
        \multirow{12}{*}{\makecell{N\textsuperscript{\underline{o}}1}\ignore{22.09.2020}} & Tx1 * & \multirow{6}{*}{Rx1} & Q-LOS & 3.5 & 4.7 & 50.7 & 21.8 & 52.9 & 9.8\\
        & Tx1 * &  & Q-LOS & 2 & 5.7 & 58.8 & 23.3 & 81.9 & 12.3\\
        & Tx2 * &  & Q-LOS & 3.5 & 7.7 & 58.9 & 42.4 & 177.4 & 24.5\\
        & Tx2 * &  & Q-LOS & 2 & 8 & 60.1 & 22.1 & 68.8 & 10.1\\
        & Tx7 * &  & NLOS & 3.5 & 16.6 & 67.3 & 35.1 & 151.6 & 27.9\\
        & Tx7 * &  & NLOS & 2 & 16.5 & 67.8 & 36.1 & 184.1 & 31.2\\ \cline{2-10}
        & Tx1 * & \multirow{3}{*}{Rx2} & NLOS & 3.5 & 12.1 & 67.3 & 49.7 & 215.6 & 49.2\\
        & Tx2 * &  & NLOS & 3.5 & 8 & 63.3 & 45.1 & 201 & 37.7\\
        & Tx4 * &  & LOS & 3.5 & 6.5 & 54 & 24.5 & 91.2 & 13.4\\ \cline{2-10}
        & Tx1 & \multirow{3}{*}{Rx4} & NLOS & 3.5 & 26.1 & 66.6 & 37.1 & 211.8 & 32.7\\
        & Tx2 * &  & NLOS & 3.5 & 21.4 & 65.7 & 41.5 & 207.2 & 33.6\\
        & Tx2 * &  & NLOS & 2 & 21.4 & 74.4 & 49.9 & 216.7 & 40.9\\
        \hline
        \multirow{19}{*}{\makecell{N\textsuperscript{\underline{o}}2}\ignore{23.05.2020}} & Tx1 * & \multirow{7}{*}{Rx1} & Q-LOS & 1 & 5.8 & 63.5 & 36.2 & 180 & 27\\
        & Tx2 * &  & Q-LOS & 1 & 8.1 & 64.6 & 51.8 & 196.7 & 37.1\\
        & Tx3 &  & Q-LOS & 1 & 12 & 65 & 41.4 & 191.2 & 35.9\\
        & Tx4 &  & NLOS & 1 & 16.7 & 69.2 & 44.4 & 212.9 & 38.8\\
        & Tx6 &  & NLOS & 1 & 21.3 & 67.8 & 49.2 & 210.9 & 34.1\\
        & Tx7 * &  & NLOS & 1 & 16.1 & 65.2 & 52.1 & 202.6 & 34.2\\
        & Tx5 &  & NLOS & 1 & 18 & 68.7 & 49.9 & 212.2 & 40\\ \cline{2-10}
        & Tx1 * & \multirow{6}{*}{Rx2} & NLOS & 1 & 12 & 66.7 & 47.2 & 202.8 & 38.6\\
        & Tx2 * &  & NLOS & 1 & 7.8 & 58.5 & 26 & 115 & 15.6\\
        & Tx3 &  & LOS & 1 & 4.9 & 56.6 & 30.2 & 124 & 17.6\\
        & Tx4 * &  & LOS & 1 & 6.2 & 53.5 & 24.8 & 107.7 & 12.9\\
        & Tx7 &  & LOS & 1 & 5.9 & 52.7 & 24 & 72.2 & 10.8\\
        & Tx5 &  & LOS & 1 & 7.4 & 55.8 & 27.7 & 94.5 & 14.7\\ \cline{2-10}
        & Tx1 & \multirow{6}{*}{Rx4} & NLOS & 1 & 26.1 & 71.6 & 45.8 & 215.1 & 40.2\\
        & Tx2 * &  & NLOS & 1 & 21.3 & 70.3 & 42.7 & 215.3 & 39.3\\
        & Tx3 &  & NLOS & 1 & 16.8 & 66.4 & 45.1 & 201.8 & 30.9\\
        & Tx4 &  & Q-LOS & 1 & 12.4 & 63.7 & 27.3 & 201.7 & 23.9\\
        & Tx7 &  & Q-LOS & 1 & 7.8 & 58.4 & 30.2 & 117.6 & 18.5\\
        & Tx5 &  & Q-LOS & 1 & 10.7 & 61.3 & 29.8 & 116.6 & 17.7\\
        \hline
    \end{tabular}
    \label{tab:summary2}

\end{table*}{}


\end{appendices}

\end{document}